\documentclass[%
 aip,
 jap,%
 amsmath,amssymb,
 reprint,%
]{revtex4-1}

\usepackage{graphicx}
\usepackage{dcolumn}
\usepackage{bm}

\begin{document}

\preprint{AIP/123-QED}

\title[Sample title]{A programmable quantum current standard from the Josephson and the quantum Hall effects}

\affiliation{Quantum metrology group, Laboratoire National de m\'{e}trologie et d'Essais, 29 avenue Roger Hennequin, 78197 Trappes, France}
\author{W. Poirier}
\email{wilfrid.poirier@lne.fr}
\author{F. Lafont}
\author{S. Djordjevic}
\author{F. Schopfer}
\author{L. Devoille}

\date{\today}

\begin{abstract}
We propose a way to realize a programmable quantum current standard (PQCS) from the Josephson voltage standard and the
quantum Hall resistance standard (QHR) exploiting the multiple connection technique provided by the quantum Hall effect (QHE) and the exactness of
the cryogenic current comparator. The PQCS could lead to breakthroughs in electrical metrology like the realization of a programmable quantum current source, a quantum ampere-meter and a simplified closure of the \emph{quantum metrological triangle}. Moreover, very accurate universality tests of the QHE could be performed by comparing PQCS based on different QHRs.
\end{abstract}

\maketitle

\section{\label{sec:level1}Introduction}

The quantum Hall effect (QHE)\cite{Klitzing1980} and the Josephson effect (JE)\cite{Josepshon1962,Shapiro1963} provide universal and reproducible resistance\cite{Jeckelmann2001,Poirier2009} and voltage\cite{Jeanneret2009} standards respectively linked to the electron charge \emph{e} and the Planck constant \emph{h} only. The QHE manifests itself by the Hall resistance quantization in a two-dimensional electron gas (2DEG) at $R_\mathrm{K}/i$ values in the non-dissipative transport limit ($i$ is an integer and $R_\mathrm{K}$ is the von Klitzing constant equal to $h/e^2$ in theory). The reproducibility of the quantum Hall resistance was checked\cite{Jeckelmann2001,Janssen2011,Schopfer2013} with a relative uncertainty of a few $10^{-11}$. Thereby, the unit ohm can be currently maintained in national metrology institutes (NMIs) with a relative uncertainty of $1 \times 10^{-9}$.
The application of the QHE in metrology was more recently enlarged by its implementation in the alternating current (AC) regime\cite{Ahlers2009}
and by the development of quantized Hall array resistance standards (QHARS)\cite{Poirier2002,Poirier2009}. These successes rely on a
technique of multiple connection which cancels the effect of any impedance connected in series to the Hall bar terminals\cite{Delahaye1993}.
Progress in resistance metrology has also resulted from the advent of resistance bridges based on cryogenic current comparators (CCC)\cite{Harvey1972}. The CCC is a perfect transformer which can measure a current ratio in terms of the number of winding turns ratio with
a relative uncertainty\cite{Poirier2009} as low as $10^{-11}$. Its accuracy relies on a flux density conservation property of the superconductive toroidal shield (Meissner effect) in which superconducting windings are embedded. Owing to a flux detector based on a superconducting quantum interference device (SQUID), a CCC can have a current sensitivity as low as 1 fA/$\sqrt{\mathrm{Hz}}$ at frequencies down to 1 Hz. The JE provides a
quantized reference voltage\cite{Jeanneret2009} $V_\mathrm{J}=n_\mathrm{J}K_\mathrm{J}^{-1}f_\mathrm{J}$ where $n_\mathrm{J}$ is an integer, $f_\mathrm{J}$ is the frequency of an electromagnetic radiation and $K_\mathrm{J}$ the Josephson constant, theoretically equal to $2e/h$. The universal character of $K_\mathrm{J}$ was tested\cite{Tsai1983,Jain1987} with an uncertainty of a few $10^{-17}$. Owing to the development of arrays composed of a large number of underdamped Josephson junctions (JJ) generating a quantized
voltage as large as 10 V, the JE is currently used in NMIs to maintain the unit volt with an accuracy of a few $10^{-10}$. Programmable Josephson array voltage
standards (PJAVS) were more recently developed to generate arbitrary voltage waveforms. They are based on arrays of overdamped JJ\cite{Kohlmann2011short} (up to 70000) divided into segments containing numbers of JJ following a binary sequence and which are independently biased by a digitally programmable current source.

Despite the unit ampere is one of the seven base units of the \emph{Syst\`eme International} (SI)\cite{BIPM}, its old definition\cite{Commentampere} cannot be implemented with the accuracy required for calibration needs. Moreover, the current unit cannot be maintained using current generators mainly because of their lack of stability and of a certain sensitivity on the load of the device under test. Nowadays, the current is rather calibrated by applying the Ohm's law to a voltage and a resistance independently calibrated in terms of $K_\mathrm{J}$ and $R_\mathrm{K}$ respectively. The relative uncertainty proposed by NMIs for current in the range from $1~\mu$A up to $100~\mu$A is not less than about $10^{-6}$. A true current standard accurate within a few $10^{-9}$ is thus missing.

To overcome this lack, a lot of research efforts have been spent to develop a quantum current standard over the last two decades\cite{Feltin2009}. The most advanced idea is based on handling of single charge carriers one by one in mesoscopic system where the charge quantization manifests itself due to Coulomb blockade\cite{Grabert1992}. Hence single-electron pumps\cite{Pothier1992} deliver a quantized current given by $I_\mathrm{Pump}=n_{Q}\times Q\times f_\mathrm{Pump}$ where $n_{Q}$, $Q$ and $f_\mathrm{Pump}$ are the number of charges per cycle, the elementary charge and the driven frequency respectively. Several kinds of single-electron pumps were proposed as reviewed in reference\cite{Pekola2013,Scherer2012}. In a seven junctions metallic pump, the reproducibility of the charge transfer has been demonstrated\cite{Keller1996} by measuring an error rate per cycle as low as $10^{-8}$. An interesting and stringent test of the accuracy of these single-electron pumps relies on a calibration of the quantized current in terms of $R_\mathrm{K}$ and $K_\mathrm{J}$, also referred to as the closure of the quantum metrological triangle (QMT)\cite{Likharev1985}. This experiment allows the measurement of the product $R_\mathrm{K}K_\mathrm{J}Q$ which is expected to be equal to 2. Any deviation from this value would question the expected theoretical values $h/e^2$, $2e/h$ and $e$ of the constants $R_\mathrm{K}$, $K_\mathrm{J}$ and $Q$ respectively\cite{Piquemal2000}. At the time of writing, the relative uncertainty of this determination\cite{Pekola2013,Scherer2012} is limited to about $10^{-6}$. Reducing this uncertainty to a few $10^{-8}$ is an issue to demonstrate the consistency of the solid state theory with an accuracy allowing an evolution towards a SI based on fixed fundamental constants\cite{BIPM}. Let us remark that closing the metrological triangle also leads to an estimate value of the elementary charge from the determinations of $R_\mathrm{K}$ and $K_\mathrm{J}$.

In this paper, we propose a way to realize a programmable quantum current standard (PQCS) from a special series arrangement of the Josephson voltage standard (JVS) and the quantum Hall resistance standard (QHR) with the CCC. We then discuss how this PQCS can lead to a breakthrough in some key metrological experiments and in particular in the closure of the QMT.

\section{\label{sec:level1}Principle of the programmable quantum current standard}

\begin{figure}[h]
\begin{center}
\includegraphics[width=8cm]{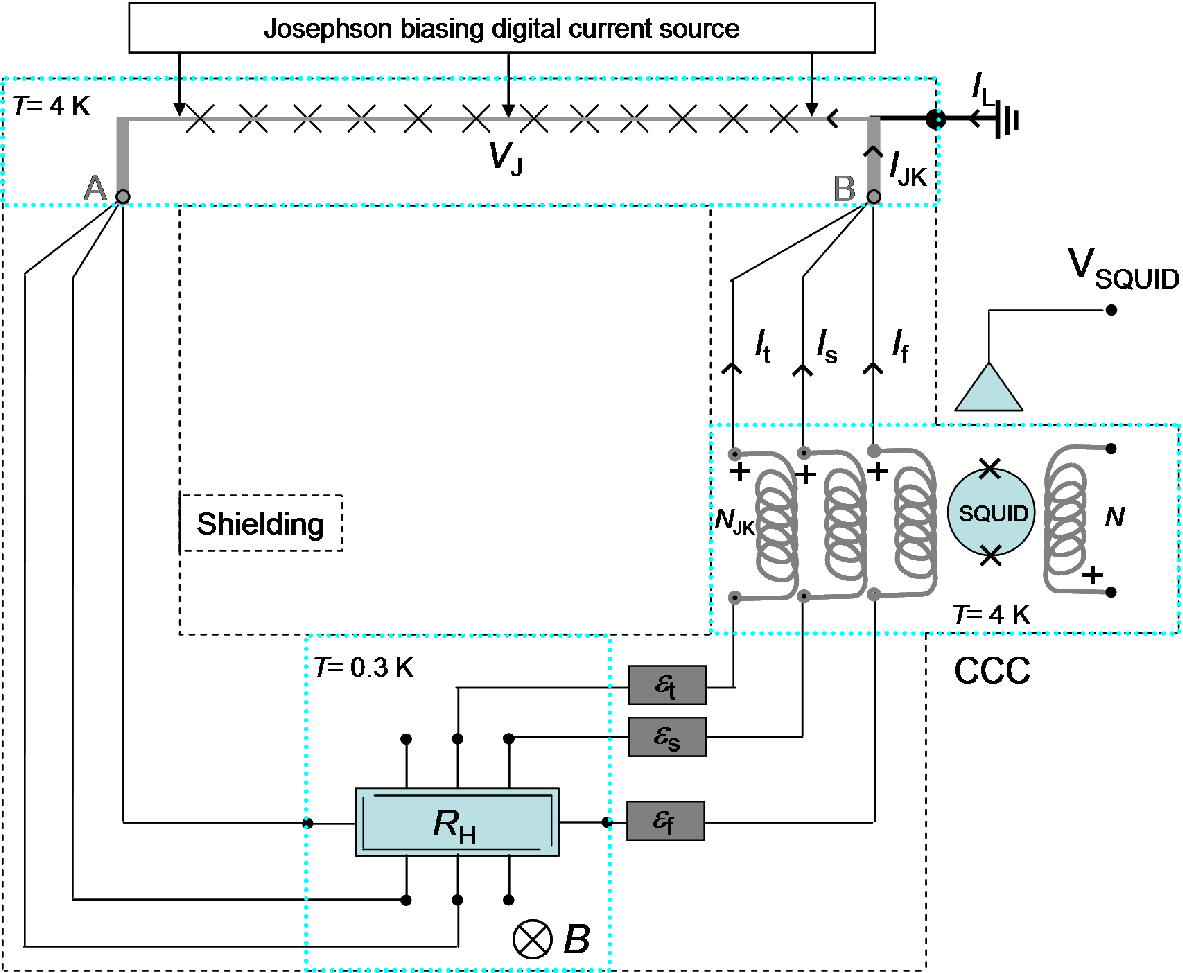}
\caption{Realisation of a PQCS from the JE and the QHE using a CCC.}\label{fig1}
\end{center}
\end{figure}
The principle of the PQCS consists in realizing the current $V_\mathrm{J}/R_\mathrm{H}$ where $V_\mathrm{J}$ is the voltage generated by a PJAVS biased on a quantized voltage step and $R_\mathrm{H}$ is the resistance of a quantum Hall resistance standard (QHR). The PQCS is theoretically quantized at current values $in_\mathrm{J}f_\mathrm{J}/(R_\mathrm{K}K_\mathrm{J})$. The main concern of the metrologist is to generate this current and then to use it as a reference for applications without degrading its quantum accuracy. The implementation is described in fig. 1. The PJAVS maintains the voltage $V_\mathrm{J}$ between the nodes A and B. The QHR of resistance $R_\mathrm{H}$ is connected at these nodes using a triple series connection (a similar connection scheme is used in AC regime\cite{Delahaye1995,Ahlers2009}) with the peculiarity that a CCC winding of number of turns $N_\mathrm{JK}$ is inserted in each of the three wires connecting the node B to the terminals of the Hall bar. A CCC auxiliary winding of $N$ turns is connected to different external circuits. Thus, the CCC allows the use of the quantized current as a reference in different metrological applications as explained in the following. Because nodes A and B which are the pads on the Josephson array chip are superconducting, the total current $I_\mathrm{JK}$ supplied by the PJAVS does not generate additional voltage drop at the nodes where the QHR is connected. Moreover, the triple series connection drastically reduces the discrepancy, caused by contact resistances, of the two-terminal resistance $R_\mathrm{AB}$ defined between A and B points to the quantized resistance $R_\mathrm{H}$. We therefore expect $I_\mathrm{JK}$ accurately close to $V_\mathrm{J}/R_\mathrm{H}$.
\begin{figure}[h]
\begin{center}
\includegraphics[width=8cm]{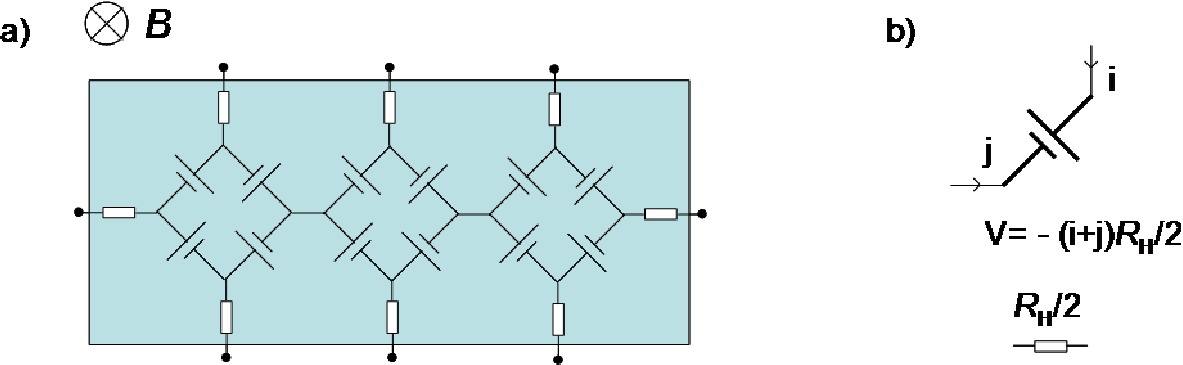}
\caption{a) Electrical model of the Hall bar based on a combination of voltage generators and resistors. b) Expression of the voltage generator as a function of the entering currents \emph{i}, \emph{j} and the Hall resistance $R_\mathrm{H}$ for the magnetic direction indicated in a) and value of the resistors ($R_\mathrm{H}/2$). }\label{fig2}
\end{center}
\end{figure}

More precisely, let us note $I_f$, $I_s$ and $I_t$ the three currents circulating in the wires from node B to the terminals of the QHR and $\varepsilon_f$, $\varepsilon_s$ and $\varepsilon_t$ the resistances of the connections expressed in relative value of $(1/2)R_\mathrm{H}$ (resistances of wires, windings and Hall bar terminal contacts). To simplify the discussion, only the resistances of the links connecting the QHR terminals (located on one equipotential edge) to the ground are considered. The currents $I_f$, $I_s$ and $I_t$ and the resistance $R_\mathrm{AB}$ can be calculated using a Ricketts and Kemeny electrical model of the Hall bar\cite{Ricketts1988} where the longitudinal resistivity is neglected (see fig. 2):
\begin{eqnarray*}
I_f&=&\frac{I_\mathrm{JK}}{(1+\frac{\varepsilon_f}{2(1+\varepsilon_s/2)}+\frac{\varepsilon_f\varepsilon_s}{4(1+\varepsilon_s/2)(1+\varepsilon_t/2)})},\\
I_s&=&\frac{\varepsilon_f}{2(1+\varepsilon_s/2)}I_f,~I_t=\frac{\varepsilon_s}{2(1+\varepsilon_t/2)}I_s,\\
R_{AB}&=&R_{\mathrm{H}}(1+(\frac{I_t}{I_\mathrm{JK}})\varepsilon_t/2)
\end{eqnarray*}
It results that the main current $I_\mathrm{JK}=V_\mathrm{J}/R_\mathrm{AB}$ differs from ($V_\mathrm{J}/R_\mathrm{H}$) by only a third order correction in $\varepsilon$:
\begin{equation}
I_\mathrm{JK}=(V_\mathrm{J}/R_\mathrm{H})(1-\frac{\varepsilon_f\varepsilon_s\varepsilon_t}{8} + O(\varepsilon^4))
\end{equation}
The ampere-turns injected in the CCC is $N_\mathrm{JK}I_\mathrm{JK}$ where $I_\mathrm{JK}=I_f + I_s + I_t$ is close to $V_\mathrm{J}/R_\mathrm{H}$ within a third order
correction in $\varepsilon$. If $\varepsilon_f, \varepsilon_s, \varepsilon_t <10^{-3}$, the current $I_\mathrm{JK}$ and the ampere-turns $N_\mathrm{JK}I_\mathrm{JK}$ depart from their quantized values ($V_\mathrm{J}/R_\mathrm{H}$) and $N_\mathrm{JK}(V_\mathrm{J}/R_\mathrm{H})$ by less than $10^{-9}$ in relative value. Thus, the proposed new connection scheme allows the exact injection of ampere-turns generated by the programmable quantized current in the CCC.

Several conditions have to be fulfilled to ensure that this ideal description holds in an experiment. Electromotive forces (emfs) generated along wires between helium temperature and room temperature should be compensated by an efficient offset substraction procedure. Increasing both the Josephson voltage and the resistance of the QHR and keeping constant their ratio reduces the impact of emfs. The QHR can be either a single Hall bar or a QHARS provided that the current supplied in the circuit is lower that the current width of the Josephson voltage steps which amounts to some mA in modern PJAVS. For a Josephson voltage of 1 V and a QHR of resistance $R_\mathrm{K}/2$ (a single Hall bar operating on the $\nu=2$ plateau where $\nu$ is the Landau level filling factor) the current is about $77~\mu \mathrm{A}$. Moreover, leakage currents responsible for a difference between the total currents circulating through the QHR and the CCC should be cancelled. These leakage currents can be strongly reduced by connecting to ground both the low potential of the Josephson array and the shielding of each wire of the setup. By this way, any leakage current $I_\mathrm{L}$ is redirected to ground. The direct leakage current short-circuiting the QHR, the most troublesome, is thus cancelled. However, the potential $\epsilon_f (R_\mathrm{H}/2)\times I_f$ at the current terminal of the QHR (potentials at the voltage terminals of the QHR are very close to ground potential) leads to the circulation of a residual small leakage current $\epsilon_f (R_\mathrm{H}/2)\times I_f/R_\mathrm{L}$ towards ground that short-circuits the first CCC winding, where $R_\mathrm{L}$ is the insulating resistance to ground of the cable. This results in the injection of an ampere-turns value in the CCC that departs from its nominal value $N_\mathrm{JK}I_\mathrm{JK}$ by about $\epsilon_f\times R_\mathrm{H}/2R_\mathrm{L}$ in relative value. Assuming $R_\mathrm{H}=R_\mathrm{K}/2$, $R_\mathrm{L}\gg 10^{12}~\Omega$ (easily achieved using cables with polytetrafluoroetylene insulator) and $\epsilon_f < 10^{-3}$ leads to a relative discrepancy much smaller than $10^{-11}$. Now we will detail some metrological applications in which the quantized current $I_\mathrm{JK}$ can be used.

\section{\label{sec:level1}Applications of the PQCS}

\subsection{\label{sec:level2}Realization of a programmable quantum current source}

A programmable quantum current source $I=(N_\mathrm{JK}/N)(V_\mathrm{J}/R_\mathrm{H})$ can be realized by servo-controlling an external current source by the PQCS. This external current source which is connected in series with a winding of $N$ turns of the CCC is controlled by the feedback signal $V_\mathrm{SQUID}$ of the SQUID electronics to null the ampere-turns value in the CCC (see Fig. 3a). The Type B\cite{GUM} evaluation of the uncertainty component of the quantum current source essentially reduces to the negligible contribution of the winding turn ratio $(N_\mathrm{JK}/N)$ uncertainty since leakage current contribution of the PQCS is strongly cancelled. The noise spectral density $S_I$ (in A/$\sqrt{\mathrm{Hz}}$) of the quantum current source caused by the Johnson-Nyquist noise of the quantum resistance standard and the SQUID noise is given by:
\begin{equation}
S_I=(1/N)\sqrt{4k_\mathrm{B}TN_\mathrm{JK}^2/R_\mathrm{H}+(G_\mathrm{CCC}S_\mathrm{SQUID})^2}
\end{equation}
$k_\mathrm{B}$ is the Boltzmann's constant, $G_\mathrm{CCC}$ is the gain of the CCC expressed in A.turns/$\phi_0$ ($\phi_0=h/e$ is the flux quantum)
and $S_\mathrm{SQUID}$ is the SQUID noise spectral density in $\phi_0/\sqrt{\mathrm{Hz}}$ and $T$ is the temperature of the QHR. Let us consider the following values: $R_\mathrm{H}=R_\mathrm{K}/2$, $G_\mathrm{CCC}=5~\mu \mathrm{A.turns}/\phi_0$, $S_\mathrm{SQUID}=3~ \mu\phi_0/\sqrt{\mathrm{Hz}}$, $T=0.3~\mathrm{K}$. Table 1 shows the excellent noise performance of the source and the expected Type A\cite{GUM} evaluation of the relative uncertainty component $U_\mathrm{A-100s}(I)/I$ of the average value over 100 secondes in the current range from tens of pA to a few mA.
\begin{table}
\caption{\label{tab:table 1} Noise and Type A uncertainty of the programmable quantum current source.}
\begin{ruledtabular}
\begin{tabular}{cccccc}
\mbox{$V_\mathrm{J}$}&\mbox{$N$}&\mbox{$N_\mathrm{JK}$}&\mbox{$I$}&\mbox{$S_I/I$}&\mbox{$U_\mathrm{A-100s}(I)/I$}\\
\hline
1 V&1&100&$\sim 7.75~$mA&$1.9\times10^{-9}/\sqrt{Hz}$&$1.9\times10^{-10}$\\
1 V&1&1&$\sim 77.5~\mu$A&$1.9\times10^{-7}/\sqrt{Hz}$&$1.9\times10^{-8}$\\
1 V&10000&1&$\sim 7.75~nA$&$1.9\times10^{-7}/\sqrt{Hz}$&$1.9\times10^{-8}$\\
10 mV&10000&1&$\sim 77.5~pA$&$1.9\times10^{-5}/\sqrt{Hz}$&$1.9\times10^{-6}$\\
\end{tabular}
\end{ruledtabular}
\end{table}

\begin{figure}[h]
\begin{center}
\includegraphics[width=8cm]{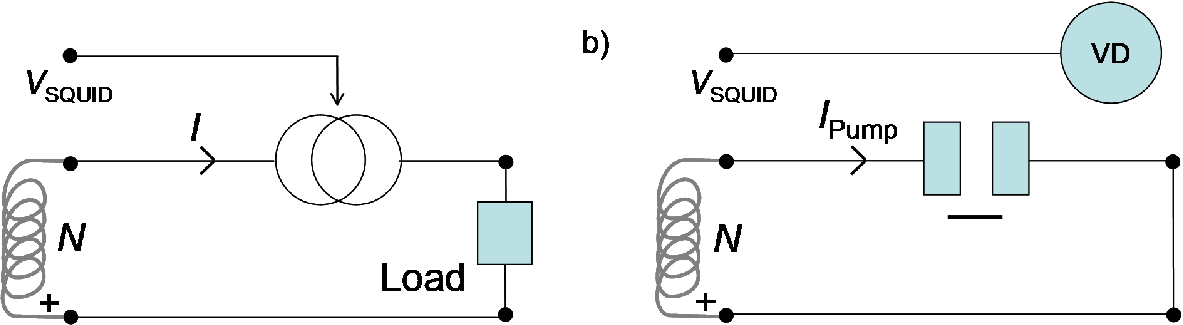}
\caption{External circuit of the CCC. a) Realisation of a programmable quantum current source. b) Realization of a quantum ampere-meter. Application to the closure of the QMT by calibrating a single-electron current pump. The SQUID output $V_\mathrm{SQUID}$ is recorded by a voltage detector VD.}\label{fig3}
\end{center}
\end{figure}

\subsection{\label{sec:level2}Realization of a programmable quantum ampere-meter to close the QMT}

Another application is the realization of a quantum ampere-meter that can be used to close the QMT. One approach to realize this experiment, proposed by the Finnish NMI (Centre for Metrology
and Accreditation)\cite{Manninen2008,Kemppinen2010}, consists in comparing the single electron current pump with the current delivered by a single Josephson junction in series with a high value resistor (typically 1 M$\Omega$) by means of a SQUID-based current null detector to measure the unbalance current. On one hand, this mounting (see fig. 4a) offers the advantage of a current detection near equilibrium, but on the other hand it faces the difficult calibration of a high value resistor placed at low temperature. A similar experiment was realized by the NMI of United Kingdom (National Physical Laboratory) using a voltage source, a current detector and a $1~\mathrm{G}\Omega$ resistor implemented at room temperature. The measurement accuracy of the current pump was limited to $1.2\times10^{-6}$ in relative value by the Johnson-Nyquist noise of the resistor and the calibration uncertainty of the room temperature apparatus\cite{Giblin2012}. The French NMI (Laboratoire National de m\'etrologie et d'Essais)\cite{Devoille2012} rather proposed to oppose the voltage of a Josephson array to the voltage drop at the terminals of a 10-k$\Omega$ resistor fed by the current obtained after amplification by the CCC of the pump current (see fig. 4b). The calibration of the resistor is greatly simplified, but this simplification is done at the expense of a more complicated detection phase. Indeed, two different null detectors are needed: a zero flux detection in the SQUID loop and a zero voltage detection in the opposition circuit at room temperature. Moreover, the current amplification relies on a SQUID feedback control of the current source supplying the resistor in series with a CCC winding which is more delicate to operate and generally causes noise enhancement. More precisely, the frequency bandwidth of the external feedback is limited by the CCC internal resonance (at a few kHz) which makes the SQUID submitted to high-frequency flux noise. Due to the intrinsic nonlinearity of the SQUID response, this can lead to unwanted rectified signals altering the measurement accuracy.

\begin{figure}[h]
\begin{center}
\includegraphics[width=8cm]{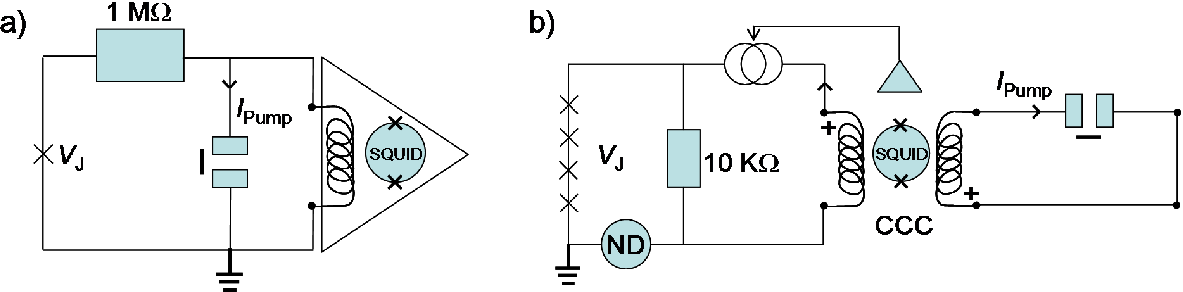}
\caption{Closing the QMT: a) by measurement of an unbalance current. b) by measurement of an unbalance voltage after amplification of the pump current. In both cases, the used resistor is independently calibrated in terms of $R_\mathrm{K}$.}\label{fig4}
\end{center}
\end{figure}
We propose, as described in Fig. 3b, to oppose by means of the CCC the ampere-turns $N_\mathrm{JK}I_\mathrm{JK}=N_\mathrm{JK}(V_\mathrm{J}/R_\mathrm{H})$
delivered by the PQCS described above to the ampere-turns $NI_\mathrm{Pump}$ generated by the
single-electron pump where $N$ is the large number of turns of a CCC winding. At equilibrium, the product $R_\mathrm{K}K_\mathrm{J}Q$ is simply determined from $(in_\mathrm{J}/n_Q)(N_\mathrm{JK}/N)(f_\mathrm{J}/f_\mathrm{Pump})$. In practice, $n_\mathrm{J}$, $f_\mathrm{J}$ and $f_\mathrm{Pump}$ can be adjusted so that the CCC detects zero ampere-turns value. The cancellation of the flux in the SQUID loop is ensured by an internal feedback. This implementation of the QMT is therefore strongly simplified since neither a voltage detector nor the implementation of an external feedback control of an amplified current source by the SQUID is required. It gathers the advantages of the two experiments described in fig. 4. The best measurement performance requires placing the single-electron current pump and the CCC in the same cryostat in order to avoid additional external noise. For $N=20000$, $N_\mathrm{JK}=1$ and the CCC characteristics considered previously, the PQCS has a contribution to the measurement noise (due to the Johnson-Nyquist noise of the QHR) one thousand lower than that of the SQUID. The relative uncertainty per $\sqrt{\mathrm{Hz}}$ of the determination of $R_\mathrm{K}K_\mathrm{J}Q$ thus boils down to $(G_\mathrm{CCC}S_\mathrm{SQUID}/N)/I_\mathrm{Pump}$.  One finds $7.5\times10^{-6}/\sqrt{\mathrm{Hz}}$ for a quantum standard nominal current of 100 pA. The target relative uncertainty of $10^{-8}$ should therefore be reached for a reasonable acquisition time of about 150 hours.

\subsection{\label{sec:level2}Comparison of quantum Hall resistance standards}

\begin{figure}[h]
\begin{center}
\includegraphics[width=8cm]{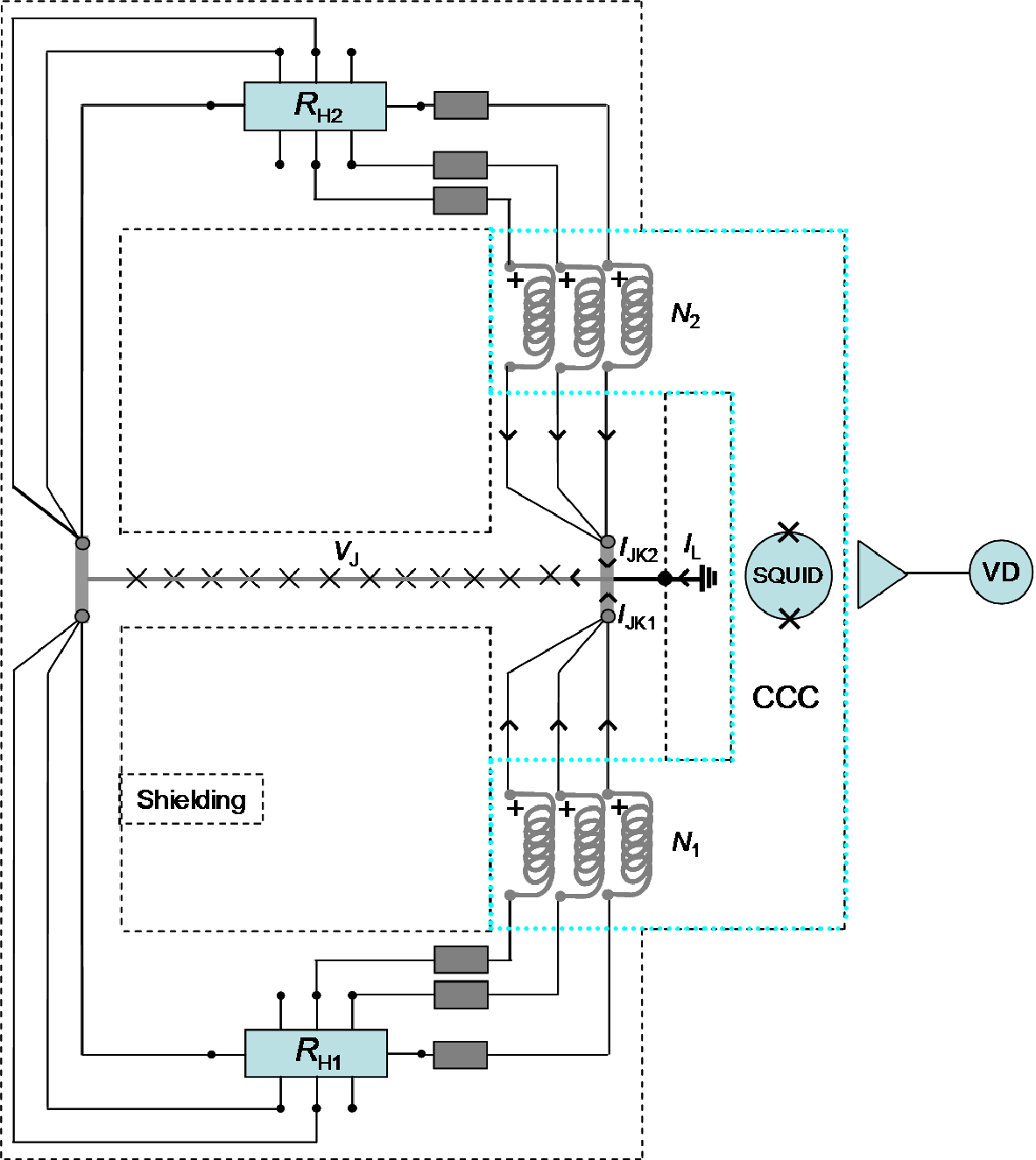}
\caption{Comparing the PQCS based on two different QHRs to realize universality tests of the QHE.}\label{fig5}
\end{center}
\end{figure}
The last application consists in comparing the resistances of two QHRs ($R_\mathrm{H1}$ and $R_\mathrm{H2}$) biased by a common Josephson voltage $V_\mathrm{J}$ (see fig. 5). Using two sets of three windings (number of turns $N_1$ and $N_2$) of the CCC, the ampere-turns $N_1I_\mathrm{JK1}$ and $N_2I_\mathrm{JK2}$ generated by the quantized currents $I_\mathrm{JK1}$ and $I_\mathrm{JK2}$ are opposed. The resistance ratio $r=R_\mathrm{H2}/R_\mathrm{H1}$ is then determined from the measurement of the ampere-turns unbalance by the CCC. In principle, a Josephson quantized voltage is not necessary in this experiment. However, the superconducting pads of the JVS ensure that the voltages biasing the two PQCS circuits are exactly the same despite the circulation of currents. Moreover, the stability and the low noise of the JVS constitute advantages to achieve the lowest measurement uncertainty for the comparison of resistances. If made of different materials, for example of graphene and GaAs, this resistance comparison leads to the realization of a very accurate universality test of the QHE. This measurement technique based on the comparison of two PQCS is more accurate than the one based on the use of a usual cryogenic resistance bridge because the measurement is not poisoned by the nanovoltmeter noise and leakage currents are efficiently cancelled. The relative uncertainty of the determination of the resistance ratio $r$ is given by:
\begin{eqnarray*}
&S&(r)/r= \\
&(&1/V_\mathrm{J})\sqrt{4k_\mathrm{B}TR_\mathrm{H1}+4k_\mathrm{B}TR_\mathrm{H2}+(R_\mathrm{H1}G_\mathrm{CCC}S_\mathrm{SQUID}/N_1)^2}
\end{eqnarray*}
Considering $N_1=N_2=2000$, $R_\mathrm{H1}=R_\mathrm{H2}=R_\mathrm{K}/2$, $V_\mathrm{J}=1~\mathrm{V}$ and $T$=0.3 K leads to $S(r)/r\simeq 0.65\times 10^{-9}/\sqrt{\mathrm{Hz}}$. This comparison method is as accurate as the Wheatstone bridge technique\cite{Schopfer2013}. Comparatively, it allows in addition the resistance comparison of quantum Hall resistors having different values. It should allow a strong improvement of the universality test of the QHE at different filling factors not only in the integer regime but also in the fractional regime.

\section{\label{sec:level1}Conclusion}

To conclude, this paper describes a PQCS realized from the Josephson and the quantum Hall effects with an accuracy limited by the Johnson-Nyquist noise of the QHR and the SQUID noise only. We show how it can be used to realize a programmable quantum current source and a quantum ampere-meter, to improve the closure of the QMT and the accuracy of quantum resistor comparisons. Let us finally remark that the PQCS could be adapted to operate in the audio-frequency (a few kHz) AC regime. This requires the implementation of a coaxial circuitry, a PJAVS high-speed biasing current source and the replacement of the CCC by an accurate AC transformer.

\begin{acknowledgments}
We wish to acknowledge O. Th\'evenot and P. Gournay as well as the staff of the LNE electrical calibration department for fruitful discussions.
\end{acknowledgments}

\providecommand{\noopsort}[1]{}\providecommand{\singleletter}[1]{#1}%

\end{document}